\documentclass[aps,prl,twocolumn,groupedaddress]{revtex4}

\usepackage{epsfig}
\usepackage{graphicx}
\usepackage{float}
\usepackage{dcolumn}
\usepackage{amsmath}
\input{epsf}

\newcommand{\ket}[1]{\left| {#1} \right\rangle}
\newcommand{\bra}[1]{\left\langle {#1}\right |}

\begin{document}

\title{Entanglement formation and violation of Bell's inequality with a semiconductor single photon source.}


\author{David Fattal}
\author{Kyo Inoue}
\altaffiliation[Also with: ]{NTT Basic Research Laboratories,
Atsugishi, Kanagawa, Japan.}

\author{Jelena Vu\v{c}kovi\'{c}}
\author{Charles Santori}
\altaffiliation[Also with: ]{Institute of Industrial Science,
University of Tokyo, Komaba, Meguro-ku, Tokyo, Japan.}

\author{Glenn S. Solomon}
\altaffiliation[Also with: ]{Solid State Photonics Laboratory,
Stanford University, Stanford, CA 94305}
 \author{Yoshihisa Yamamoto}
\altaffiliation[Also with: ]{NTT Basic Research Laboratories,
Atsugishi, Kanagawa, Japan.}


\affiliation{Quantum Entanglement Project, ICORP, JST\\ Ginzton
Laboratory, Stanford University, Stanford CA 94305}


\date{\today}

\begin{abstract}

We report the generation of polarization-entangled photons, using
a quantum dot single photon source, linear optics and
photodetectors. Two photons created independently are observed to
violate Bell's inequality. The density matrix describing the
polarization state of the postselected photon pairs is also
reconstructed, and agrees well with a simple model predicting the
quality of entanglement from the known parameters of the single
photon source. Our scheme provides a method to generate no more
than one entangled photon pair per cycle, a feature useful to
enhance quantum cryptography protocols using entangled photons.

\end{abstract}


\maketitle


Entanglement, the counter-intuitive non-local correlations allowed
by quantum mechanics between distinct systems, has recently drawn
much attention due to its applications to the manipulation of
quantum information \cite{ref:ent_review}. These non-local
correlations are often understood as the result of prior
interactions between the quantum mechanical systems of interest.
Following this idea, and as often quoted, entanglement would
represent the \textit{memory} of those interactions. But as
sugested by the Innsbruck teleportation experiment
\cite{ref:Innsbruck}, this is too limited a view. Entanglement can
be induced between completely independent particles, due to the
lack of which-path information, or in other words to the quantum
indistinguishability of two identical particles. Pionneering work
by Shih and Alley \cite{ref:Shih}, followed by Ou and Mandel
\cite{ref:Mandel2}, already used the quantum indistinguishability
to induce entanglement between two photons produced in a
non-linear crystal. More recently, entanglement swapping
experiments \cite{ref:swapping,ref:GHZ} used two independent
entangled photon pairs to induce entanglement between photons of
different pairs which never interacted. In this paper, we use a
similar linear optics technique to induce polarization
entanglement between single photons emitted independently in a
semiconductor quantum dot source, at 2 ns time interval. We
observed a clear violation of Bell's inequality, which constitutes
an experimental proof of non-local behavior for the first time
with a semiconductor single photon source. The complete density
matrix describing the polarization state of the two photons was
also reconstructed, and satisfies the Peres criterion for
entanglement \cite{ref:Peres}. We show that our results can be
quantitatively explained in terms of basic parameters of the
single photon source and derive a simple criterion for
entanglement generation using those parameters. Eventually, we
explain why our technique can be applied to quantum key
distribution (QKD) in a straightforward and useful manner.

This experiment relies on two crucial features of our quantum dot
single photon source, namely its ability to suppress multi-photon
pulses \cite{ref:Charlie01}, and its ability to generate
consecutively two photons that are quantum mechanically
indistinguishable \cite{ref:Charlie02}. The idea is to "collide"
these photons with orthogonal polarizations at two conjugated
input ports of a non-polarizing beam splitter (NPBS). A quantum
interference effect ensures that photons simultaneously detected
at different output ports of the NPBS should be entangled in
polarization \cite{ref:Mandel2}. More precisely, when the two
optical modes corresponding to the output ports 'c' and 'd' of the
NPBS have a simultaneous single occupation, their joint
polarization state is expected to be the EPR-Bell state:
\[\ket{\Psi^-} = \frac{1}{\sqrt{2}}(
\ket{H}_c\ket{V}_d-\ket{V}_c\ket{H}_d)\] Denoting 'a' and 'b' the
input port modes of the NPBS, they are related to the output modes
'c' and 'd' by the 50-50\% NPBS unitary matrix according to:
\[ a_{H/V} = \frac{1}{\sqrt{2}}(c_{H/V}+d_{H/V})\]
\[ b_{H/V} = \frac{1}{\sqrt{2}}(c_{H/V}-d_{H/V})\] where subscripts
'H' and 'V' specify the polarization (horizontal or vertical) of a
given spatial mode. The quantum state corresponding to single-mode
photons with orthogonal polarizations at port 'a' and 'b' can be
written as: \[ a^{\dag}_Hb^{\dag}_V\ket{vac} =
\frac{1}{2}(c^{\dag}_Hc^{\dag}_V - d^{\dag}_Hd^{\dag}_V -
c^{\dag}_Hd^{\dag}_V + c^{\dag}_Vd^{\dag}_H) \ket{vac}\] It is
interesting to note that this state already features non-local
correlations and violates Bell's inequality, as pointed out in
\cite{ref:Popescu97}. In this sense the post-selection is not an
essential part of entanglement formation from two identical
quantum particles. The experimental violation of Bell's inequality
with the global state would however require the discrimination
between one-photon and two-photon pulses. Instead, if we discard
the events when two photons occupy the same spatial mode 'c' or
'd' at the output (which is done naturally by detecting
coincidence counts between 'c' and 'd'), we obtain the
postselected state:
\[ \frac{1}{\sqrt{2}}(c^{\dag}_Hd^{\dag}_V -
c^{\dag}_Vd^{\dag}_H)\ket{vac} = \ket{\Psi^-} \] as claimed. The
post-selection can be done with regular single photon counter
modules. Note that the generation of polarization entangled states
via two-photon cascade emission \cite{ref:Aspect} and parametric
down converter \cite{ref:PDC_ent}
also rely upon a photon number post-selection.\\

\begin{figure}[t]
\begin{center}
   \epsfig{file=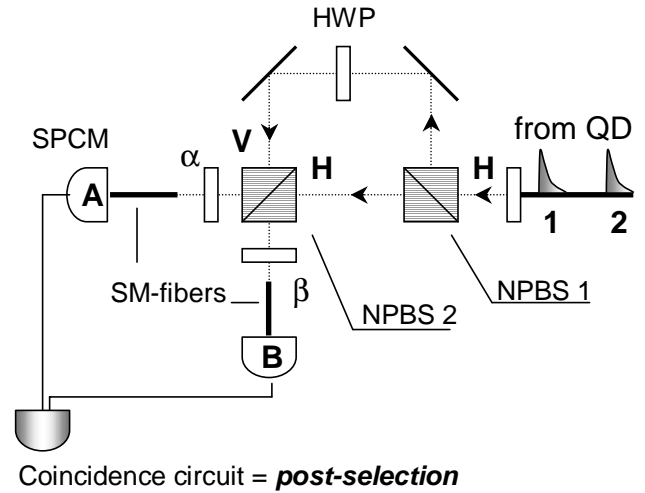, bbllx=150, bblly=95, bburx=405, bbury=425, width=3in, angle=-90, clip=}

    \caption{Experimental setup. Single photons from the QD microcavity device are sent through a single mode fiber, and have their polarization rotated to
    H. They are split by a first NPBS. The polarization is changed to V in the longer arm of the Mach-Zender configuration.
    The two path of the interferometer merge at a second NPBS. The output modes are matched to single mode fibers for subsequent detection.
    The detectors are linked to a time-to-amplitude converter for a record of coincidence counts.}
    \label{fig:setup}
\end{center}
\end{figure}

The experimental setup is shown in fig \ref{fig:setup}. The single
photon source consists of a self-assembled InAs quantum dot (QD)
embedded in a GaAs/AlAs DBR microcavity \cite{ref:Charlie02}. It
was placed in a Helium flow cryostat and cooled down to 4-10 K.
The temperature was adjusted to tune the QD emission wavelength to
cavity resonance. This increases the source brightness and reduces
the effects of dephasing by increasing the radiative decay rate
\cite{ref:Gerard01,ref:YYpaper2}. Single photon emission was
triggered by optical excitation of a single QD, isolated in a
micropillar. We used 3 ps Ti:Sa laser pulses on resonance with an
excited state of the QD, insuring fast creation of an
electron-hole pair directly inside the QD. Pulses came by pairs
separated by 2 ns, with a repetition rate of 1 pair/13 ns. The
emitted photons were collected by a single mode fiber and sent to
a Mach-Zender type setup with 2 ns delay on the longer arm. A
quarter wave plate (QWP) followed by a half wave plate (HWP) were
used to set the polarization of the photons after the input fiber
to linear and horizontal. An extra half wave plate was inserted in
the longer arm of the interferometer to rotate the polarization to
vertical. One time out of four, the first emitted photon takes the
long path while the second photon takes the short path, in which
case their wavefunctions overlap at the second non-polarizing
beam-splitter (NPBS 2). In all other cases, the single photon
pulses "miss" each other by at least 2 ns which is greater than
their width (100 - 200 ps). Two single photon counter modules
(SPCMs) in a start-stop configuration were used to record
coincidence counts between the two output ports of NPBS 2,
effectively implementing the post-selection (if photons exit NPBS
2 by the same port, then no coincidence are recorded by the
detectors). Single-mode fibers were used prior to detection to
facilitate the spatial mode-matching requirements (they actually
\textit{define} the output modes). They were preceded by quarter
wave and polarizer plates to allow the
analysis of all possible polarizations.\\

\begin{figure}[hbtp]
    \epsfig{file=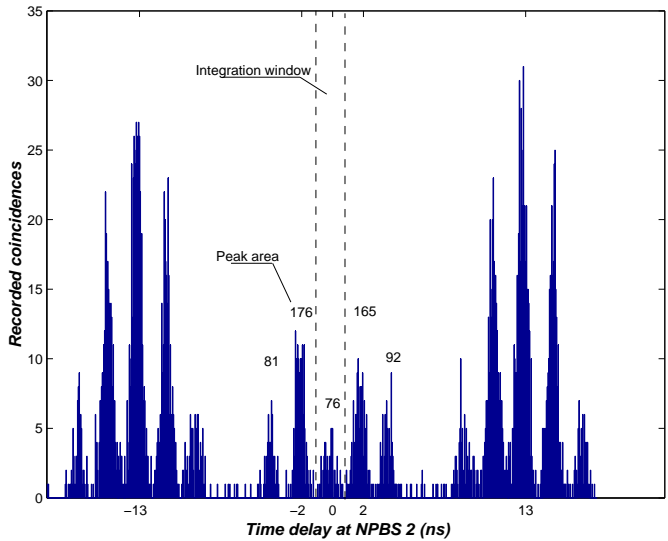, width=3.5in}
    \caption{Zoom on a typical correlation histogram, taken on $QD_1$.
    Coincidences were actually recorded over a time window of 100 ns. The integration time was 2 min, short enough to guarantee that the QD
    is illuminated by a constant pump power. The central window corresponds to photons that overlapped at the NPBS and took
    different output ports, i.e. the post-selected events.}
    \label{fig:example}
\end{figure}

The detectors were linked to a time-to-amplitude converter, which
allowed to record histograms of coincidence events versus
detection time delay $\tau$. A typical histogram is shown on fig
\ref{fig:example}. The integration time was two minutes. The
number of coincidences for overlapping photons was measured as the
area of the peak contained in the domain -1ns$<\tau<$1ns. For
given analyzer angle settings $(\alpha, \beta)$, we denote by
$C(\alpha,\beta)$ this number normalized by the total number of
coincidences in a time window of 100 ns. This normalization is
independent of $(\alpha, \beta)$ since the input of NPBS 2 are two
modes with orthogonal polarizations. $C(\alpha,\beta)$  measures
the average rate of coincidences throughout the time
of integration.\\

Two different QD microcavity devices were used to produce single
photons. Both of them featured a high suppression of two-photon
pulses and high overlap (indistinguishability) between consecutive
photons. The overlap was measured by the Mandel dip
\cite{ref:Charlie02}, which was estimated by removing the HWP in
the long arm, thus colliding completely identical particle at NPBS
2. A Mandel dip of zero means perfect indistinguishability between
consecutively emitted single photons. Interestingly, the
by-product of a Mandel test should be a photon-number entangled
state $\ket{0,2}-\ket{2,0}$. However, the coincidence measurements
alone presented in ref \cite{ref:Charlie02} do not rule out the
possibility of a decohered mixture
$\ket{0,2}\bra{0,2}+\ket{2,0}\bra{2,0}$. This decoherence issue
will be fully addressed in the present work.

 A Bell's inequality test was performed for post-selected photon
pairs from $QD_1$. Following ref \cite{ref:cshs}, if we define the
correlation function $E(\alpha,\beta)$ for polarizer angle
settings $\alpha$ and $\beta$ as:
\[E(\alpha,\beta) = \frac{C(\alpha,\beta) + C(\alpha^{\bot},\beta^{\bot}) -
C(\alpha^{\bot},\beta)- C(\alpha,\beta^{\bot})}{C(\alpha,\beta) +
C(\alpha^{\bot},\beta^{\bot}) + C(\alpha^{\bot},\beta)+
C(\alpha,\beta^{\bot})} \] then local realistic assumptions lead
to the inequality:
\[S = |E(\alpha,\beta) - E(\alpha{\prime},\beta)| + |E(\alpha{\prime},\beta{\prime}) +
E(\alpha,\beta{\prime})|\leq 2 \] that can be violated by quantum
mechanics.\\

Sixteen measurements were performed for all combination of
polarizer settings among $\alpha \in \{0^o, 45^o, 90^o, 135^o\}$
and $\beta \in \{22.5^o, 67.5^o, 112.5^o, 157.5^o\}$. The
corresponding values of normalized coincidence counts $C(\alpha,
\beta)$ are reported in table \ref{tab:bell}, up to an overall
multiplicative constant. The statistical error on $S$ is quite
large, due to the short integration time used to insure high
stability of the QD device. Bell's inequality is still violated by
two standard deviations, according to $S \sim 2.38 \pm 0.18$. This
result constitutes the first observation of non-local correlations
created between two single independent photons by linear-optics
and photon number post-selection. Entanglement was created from a
completely separable photon pair state.

\begin{table}

\begin{ruledtabular}
\begin{tabular}{c|cccc}
 $\beta\, \backslash\, \alpha$ &$0^o$ & $45^o$ & $90^o$ & $135^o$\\
 \hline\\

  $22.5^o$ & 8.2  &  41.8 &  42.0  &  6.9\\
  $67.5^o$ & 13.3 &  12.2 &  37.0  & 36.9\\
  $112.5^o$ & 42.5 &  7.9  &  6.7   & 41.7\\
  $157.5^o$ & 38.3 &  36.6 &  12.7  & 12.9\\
\end{tabular}
\end{ruledtabular}
\caption{\label{tab:bell}Normalized coincidences $C(\alpha,\beta)$
for different polarizer angles used in the Bell's inequality test.
The units are arbitrary. Note that $C(\alpha,\beta) +
C(\alpha^{\bot},\beta^{\bot}) + C(\alpha^{\bot},\beta)+
C(\alpha,\beta^{\bot})$ is constant $\sim 100$ for given settings
$\alpha$ and $\beta$.}
\end{table}

\begin{figure}[hbtp]
    \epsfig{file=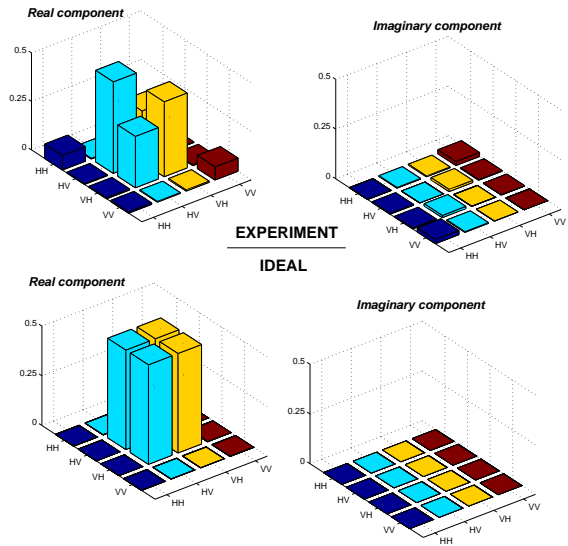, bbllx=0, bblly=79, bburx=652, bbury=741, width=3in, clip=}
    \caption{Reconstructed polarization density matrix for the post-selected photon pairs
    emitted by $QD_2$. The small diagonal HH and VV components are caused by finite
    two-photon pulses suppression ($g^{(2)} >0$). Additional reduction of the off-diagonal elements originates from the imperfect
    indistinguishability between consecutively emitted photons.}
    \label{fig:dens_mat}
\end{figure}

To understand the degree of entanglement in detail, we
reconstructed the postselected two-photon state, for comparison
with a simple model. The two-photon polarization state can be
completely characterized by a reduced density matrix, where only
the polarization degrees of freedom are kept. This density matrix
can be reconstructed from a set of 16 measurements with different
polarizer settings, including circular \cite{ref:Kwiat}. We
performed this analysis, know as \textit{quantum state
tomography}, on photon pairs emitted by $QD_2$. The reconstructed
density matrix is shown on fig \ref{fig:dens_mat}. It can be shown
to be non separable, i.e. entangled, using the Peres criterion
\cite{ref:Peres} (negativity $\sim  0.43$, where a value of 1
means maximum entanglement).

 We next try to account for the observed
degree of entanglement from the parameters of the QD single photon
source. Due to residual two-photon pulses, giving a non-zero value
to the equal time second-order correlation function $g^{(2)}(0)$
\cite{ref:Charlie01}, a recorded coincidence count can originate
from two photons of same polarization that would have entered NPBS
2 from the same port. A multi-mode analysis also reveals that an
imperfect overlap $V = \left|\int \psi_1(t)^* \psi_2(t)\right|^2$
between consecutive photon wavefunctions washes out the quantum
interference responsible for the entanglement generation.
Including those imperfections, we could derive a simple model for
the joint polarization state of the postselected photons. In the
limit of low pump level, this model predicts the following density
matrix in the (H/V)$\otimes$(H/V) basis:
\[ \rho_{model} = \frac{1}{\frac{R}{T}+\frac{T}{R}+4g^{(2)}}\left(
\begin{array}{cccc}
2g^{(2)} & & & \\
 & \frac{R}{T} & -V &  \\
 & -V & \frac{T}{R} & \\
 & & & 2g^{(2)}\\
\end{array} \right)  \] R and T are the reflection and
transmission coefficients of NPBS 2 ($\frac{R}{T} \sim 1.1$ in our
case). Using the values for $g^{(2)}$ and $V$ measured
independently, we obtain an excellent quantitative agreement of
our model to the experimental data, with a fidelity $Tr \left(
\sqrt{\rho_{exp}^{\frac{1}{2}} \, \rho_{model} \,
\rho_{exp}^{\frac{1}{2}}} \right)$ as high as 0.997.\\
The negativity of the state $\rho_{model}$ is proportional to
$(V-2g^{(2)})$, which means that entanglement exists as long as
$V>2g^{(2)}$. This simple criterion indicates whether a given
single
photon source will be able to generate entangled photons in such a scheme.\\

We now study the possible improvements and applications of our
entanglement generation. If an optical switch is used to direct
each photon on its proper path, our scheme will ideally succeed
half of the time. Moreover, post-selection implies that the
photons are destroyed when our scheme succeeds. This is a serious
obstacle for some applications to quantum information systems, but
not all. Indeed, the Ekert91 \cite{ref:Ekert} or BBM92
\cite{ref:BBM92} QKD protocols using entangled photons can
directly be performed with our technique. The essence of these
protocols is to establish a secure key upon local measurement of
two distant photons from an entangled pair, which is exactly
similar to our scheme. The bit error induced by uncorrelated
photon pairs in those protocols is significantly suppressed
\cite{ref:Edo1} when single entangled pairs are used, a feature
which only our source possesses among the currently demonstrated
entangled photon sources. Therefore, those QKD protocols should
actually benefit from our method to generate entanglement. When
the scheme fails, one party, say Alice, does not receive any
photon, so that Bob will discard his result. The intermittent
failure of the scheme will effectively halve the communication
rate, without compromising the secrecy of the key. \\

In summary, we demonstrated the violation of Bell's inequality for
the first time with a semiconductor single photon source.
Polarization entanglement was induced between two independent but
indistinguishable single photons, with linear-optics only. Our
technique naturally produces no more than one entangled pair per
cycle, which is a unique feature among previously demonstrated
entangled photon sources. Our scheme can be straightforwardly
applied to Ekert91/ BBM92 QKD, and should perform better than
current entangled photon sources for that purpose.

\end{document}